\documentclass[10pt]{wlscirep}
\usepackage[utf8]{inputenc}
\usepackage[T1]{fontenc}
\usepackage{placeins}
\usepackage{setspace}
\usepackage{lineno}
\usepackage{soul}

\newcommand{\acc}[2]{#1\,{\scriptsize$\pm$\,#2}}

\usepackage{graphicx} 
\usepackage[inkscapelatex=false]{svg}
\usepackage{pdfpages}
\usepackage{tikz}
\usepackage{amsthm}
\usepackage{amsmath}
\usepackage{bbm}
\usepackage{amssymb}
\usepackage{placeins}
\usepackage{xcolor}

\usepackage{caption}
\usepackage{subcaption}
\usepackage{algorithm}
\usepackage{algpseudocode}
\usepackage{multirow}
\usepackage[flushleft]{threeparttable}
\usepackage{bm}
\usepackage{comment}
\usepackage{hyperref}

\usepackage{pifont}

\newcounter{suppnote}
\newcommand{\fakesupp}[1]{%
    \refstepcounter{suppnote}%
    \label{#1}%
}
\fakesupp{supp:locality_criteria}  
\fakesupp{supp:spectral_energy_mnist_stl}
\fakesupp{supp:Noise_filtering}
\fakesupp{supp:SignUpdateProperty}
\fakesupp{supp:Noise_prop}
\fakesupp{supp:Gradients}
\fakesupp{supp:mlp_negative_control}
\fakesupp{supp:cifar10_dev_bs256}         
\fakesupp{supp:read_noise}                
\fakesupp{supp:batch_svhn}                  
\fakesupp{supp:HPO}

\title{Local learning with imperfect analog synapses}

\author[1,*]{Th\'eo Dupuis} 
\author[1]{Adrien Renaudineau} 
\author[3]{Guillaume Bourgeois} 
\author[3]{Joel Minguet Lopez} 
\author[3]{Gabriele Navarro} 
\author[2]{Jean-Michel Portal}
\author[3]{Elisa Vianello}
\author[1]{Jacques-Olivier Klein} 
\author[1,*]{Damien Querlioz} 
\affil[1]{Universit\'e Paris-Saclay, CNRS, Centre de Nanosciences et de Nanotechnologies, Palaiseau, France}
\affil[2]{Aix-Marseille Université, CNRS, Institut Matériaux Microélectronique Nanosciences de Provence, Marseille, France}
\affil[3]{Universit\'e Grenoble Alpes, CEA-Leti, Grenoble, France}
\affil[*]{theo.dupuis@cnrs.fr, damien.querlioz@universite-paris-saclay.fr}

\begin{abstract}
Analog in-memory computing delivers efficient neural network inference at the edge, but systems deployed in non-stationary environments may need to adapt in situ to locally acquired data. Backpropagation faces two structural obstacles here. Its global error transport lets stochastic write errors compound across layers, and its backward pass demands bidirectional, high-precision peripheral circuitry absent from inference-oriented crossbar arrays. We introduce Channel-wise Spatial Gradient learning (CwSG), a strictly local, forward-only rule that trains each convolutional layer by rewarding spatial variation in its feature maps. In simulations replaying the measured trajectories of 1{,}189 phase-change memory and 1{,}268 filamentary memristor devices, CwSG learns with single, unverified, sub-1 V pulses. It matches backpropagation on phase-change memory with 59-fold fewer pulses and outperforms it by 53 percentage points on highly variable filamentary memristors, with 8-bit forward converters and without transposed crossbar reads or network-wide activation storage. Analog training thus becomes a problem of credit assignment, in which algorithmic design rather than hardware compensation alone overcomes severe programming noise.
\end{abstract}

\begin{document}

\maketitle


\thispagestyle{empty}

\section*{Introduction}\label{main:Introduction}

Analog in-memory computing (AIMC) delivers highly energy-efficient edge inference by performing matrix-vector multiplications directly within non-volatile memory arrays, such as phase-change memory (PCM) and filamentary memristors (FMs) \cite{Hermes_pcm_inf, kim20214k, wan2022compute, ambrogio2023analog, huang2024memristor, hung2021four}. Extending the same arrays from inference to learning would let deployed edge systems adapt to locally acquired data, with potential benefits for energy, latency, and privacy \cite{kang2025bridging, winkle2020situ, hundman2018detecting}. Yet the very architecture that makes AIMC efficient for inference is difficult to reconcile with how neural networks are conventionally trained.

The first challenge is credit assignment: determining how each layer should change its weights. Standard offline training relies on backpropagation (BP), which retains intermediate activations and transports errors backward through successive transposed weight matrices \cite{Lillicrap2016-FA, akrout2019deep, woo2021activation}. Mapped onto AIMC, this global dataflow demands bidirectional array access, duplicated peripheral converters, or separately programmed copies of the transposed weights \cite{wan2022compute, Marinella18BidirCrossbar, Narayanan17ibmBidirCrossbar, Song17pipelayer, xiao2020analog}, undermining the simple, unidirectional architecture of inference-oriented crossbars.

Forward-only, layer-local learning therefore offers a natural alternative. The original Forward-Forward proposal showed that each layer can be trained on its own ``goodness'' objective, removing inter-layer backward error transport and network-wide activation retention \cite{hinton2022forward}, although the original algorithm underperforms on deep networks. Supervised local methods have since made this approach substantially more effective for deep convolutional architectures. Convolutional Channel-wise Competitive learning (CwC), in particular, associates groups of channels directly with classes: it obtains a supervised prediction from the active layer alone, without an auxiliary classifier or a second presentation of modified data
 \cite{Papachrist24ClusterF}.

The second challenge is physical: replacing BP with a layer-local objective does not, in itself, make learning robust to imperfect devices. Analog non-volatile memories exhibit device-to-device variability, state-dependent conductance evolution, and stochastic programming steps \cite{Gaba2013,Zahoor2020}. Controlling the size of a physical weight change is therefore difficult. Locality helps by removing the dependence of each layer's learning signal on imperfect weights in later layers. The choice of local objective remains critical, however: 
a local rule developed under precise numerical optimization may become unstable, drift systematically, or favor degenerate solutions once gradient magnitudes are distorted.

CwC illustrates this general problem concretely. Its goodness score rewards the activation energy of each class-associated channel group, allowing a nearly uniform map to score highly without encoding spatial detail. This offers a shortcut to high goodness: increasing activity broadly rather than learning a selective spatial response. This tendency, already visible under ideal numerical updates, is reinforced by stochastic programming, as broad increases in activity can be easier to achieve than carefully shaped spatial responses.

Here, we show that robust learning with imperfect analog synapses can emerge from changing what each layer rewards: spatial structure instead of activation amplitude. We introduce Channel-wise Spatial Gradient learning (CwSG), which preserves the features that let CwC train deep networks (class-to-channel supervision, classifier-free local prediction, and a greedy layer-wise schedule) but replaces activation-amplitude goodness with finite-difference spatial-gradient energy. A uniform feature map now scores exactly zero, regardless of its amplitude, whereas edges, contours, and textures contribute directly. Each layer is thus rewarded not simply for responding strongly, but for producing spatially differentiated responses. We show that this change does more than improve convolutional accuracy: it also supports effective learning under sign-only updates and stochastic device programming.

Under full-precision optimization, CwSG achieves the highest accuracy among the evaluated backpropagation-free methods on MNIST, Fashion-MNIST, and CIFAR-10, approaching end-to-end BP to within 0.2, 0.6, and 2.6 percentage points, respectively. When constrained to fixed-magnitude, sign-only updates, CwSG loses only 0.6, 1.5, and 0.4 percentage points on the same tasks, whereas standard BP loses 25.6 points on CIFAR-10. In controlled write-noise experiments with sign-only updates, CwSG tolerates up to an order of magnitude higher noise levels than BP.

We then evaluate CwSG in device-calibrated simulations that replay experimentally measured programming trajectories from 1,189 PCM and 1,268 FM devices. On SVHN \cite{Netzer2011}, CwSG reaches $82.5\%$ on PCM, comparable to the $82.2\%$ obtained by direct fixed-pulse in-situ BP, while requiring only 6.4 rather than 378 programming pulses per device, a 59-fold reduction. Conversely, on the quasi-linear but more variable FM trajectories, CwSG reaches $84.9\%$, whereas the evaluated BP implementation remains at $31.5\%$. CwSG also retains its accuracy with 8-bit forward converters and never requires a transposed crossbar read. Robust analog training, therefore, does not need to rely on precise device programming: it can emerge from a learning rule designed around the native operations of the hardware.

Active compensation offers another route to robust analog learning by improving the fidelity of physical weight updates. Iterative program-and-verify \cite{esmanhotto2022experimental, rao2023thousands, Hermes_pcm_inf}, mixed-precision digital accumulation \cite{LeGallo2017}, and auxiliary analog accumulators \cite{Rasch2024, Gokmen20_tikitaka, ambrogio2018equivalent} can all deliver accurate updates, but at the price of repeated reads and writes, digital accumulation, or scheduled transfer operations. CwSG targets the inexpensive programming regime in which each update is a single low-voltage soft SET or RESET pulse, applied without reading or verifying the device state \cite{li2018efficient, imbert2024forward}. Our previous experiments demonstrated supervised Forward-Forward and competitive-forward learning with such updates in FM classifier layers operating on pretrained image features \cite{renaudineau2026}. CwSG extends this approach to training the convolutional layers themselves.

The paper proceeds in three stages. We first compare CwSG with BP and other local methods under ideal numerical updates, then introduce sign-only updates and programming noise. Finally, we evaluate learning with measured device responses, finite-resolution converters, and small batches.

\section*{Results}\label{main:Results}

\subsection*{A spatial-gradient objective improves local convolutional learning}\label{main:Results:LocalvsBP}

We begin by defining the CwSG learning rule and establishing its algorithmic viability independently of analog synaptic behavior. By evaluating its performance under full-precision optimization, we isolate its capacity to solve the architectural problem of weight transport before introducing physical device non-idealities.

\begin{figure}[h!t]
    \centering
    \includesvg[inkscapelatex=false,width=\textwidth]{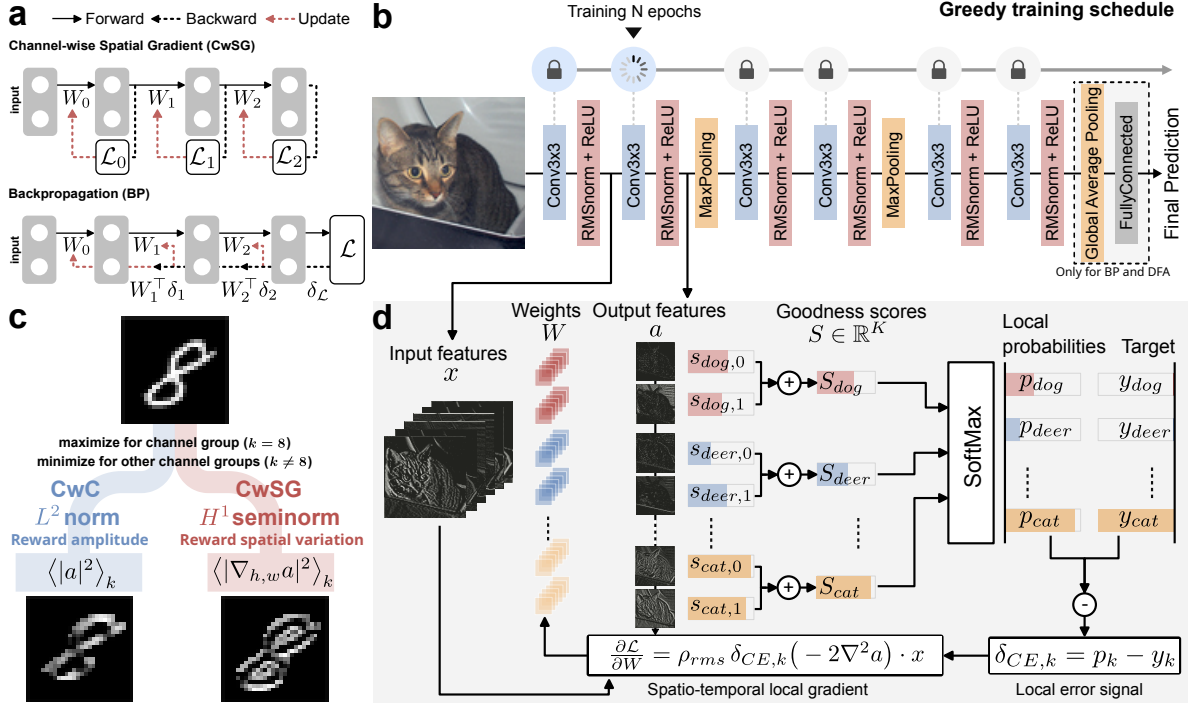}
    \caption{\textbf{Channel-wise spatial-gradient learning enables strictly local credit assignment.}
    \textbf{a} Credit assignment comparison. CwSG evaluates a supervised loss at the active layer and strictly confines the gradient computation to that layer. Backpropagation transports global errors through successive transposed-weight operations $W_l^{\top}\delta_l$.
    \textbf{b} Greedy training schedule on CNN-7. Previously trained layers act as a frozen feature extractor, the active layer optimizes its local loss, and downstream layers remain uncomputed. The RMSnorm is parameter-free. At inference, network predictions rely exclusively on the per-class goodness scores produced by the final layer.
    \textbf{c} CwC and CwSG goodness functions, illustrated by the layer activations for an MNIST digit `8' after training to maximize each respective objective. CwC rewards squared activation amplitude, whereas CwSG rewards spatial-gradient energy across the horizontal and vertical axes. A spatially uniform high-amplitude response therefore yields high CwC goodness but exactly zero CwSG goodness.
    \textbf{d} Exact computation for a single CwSG layer. The local activation error propagates backward only through the parameter-free ReLU and RMS normalization steps to form the weight update $\frac{\partial \mathcal{L}}{\partial W}$. The gradient is computed exclusively from spatially and temporally local information and does not rely on physical weight reads.
    }
    \label{fig:Fig1}
\end{figure}

We build upon the supervised architecture of CwC \cite{Papachrist24ClusterF} because it satisfies three hardware locality criteria: (i) forward-only updates without storing intermediate activations across multiple layers, (ii) no transposed backward projections or weight transport, and (iii) strictly rely on a layer-local error signal (Fig.~\ref{fig:Fig1}a). Alternative approaches address some of these constraints, but they typically retain some form of nonlocal dataflow or introduce auxiliary supervision mechanisms. Direct Feedback Alignment (DFA) methods and its variants remove exact weight symmetry but still distribute a global error signal that requires network-wide activation storage \cite{Lillicrap2016-FA, DFA, lv2025_DLL}. Other layer-wise methods derive supervision through auxiliary classifier heads or embedding comparisons, as in Dictionary Contrastive Learning (DCL) \cite{choi2024_DCL}, introducing shared external embeddings, additional trainable weights (l-DCL), or peripheral computation \cite{choi2024_DCL, nokland2019local, belilovsky2020greedy}. CwC instead  maps groups of output channels directly to target classes, deriving a supervised learning signal from the active layer's inputs, activations, and label. This classifier-free, forward-only operation eliminates the need for auxiliary heads and fulfills all three locality criteria by construction, providing an ideal foundation for CwSG. Supplementary Note~\ref{supp:locality_criteria} provides formal definitions of these criteria and a detailed comparison across the evaluated learning rules.

Layers are trained greedily: while layer $\ell$ learns, previously trained layers remain frozen and downstream layers are not evaluated (Fig.~\ref{fig:Fig1}b). At inference, CwSG predicts the class using only the goodness scores of the final layer.

For a single convolutional layer, we define the output channel activation $a_{c_{out}}$ at each spatial position $\xi=(h,w)$ by 

\begin{equation}
    a_{c_{out}}(\xi) = \text{ReLU}\left(W * x\right), 
    \label{eq:forward_pass}
\end{equation}
where $x \in \mathbb{R}^{C_{in} \times H \times W}$ is the input tensor, $W \in \mathbb{R}^{C_{out} \times C_{in} \times K_h \times K_w}$ the weight tensor, and $*$ the 2D convolution operator. The output channels $a_{c_{out}}$ are partitioned into $K$ disjoint class groups $\mathcal{C}_{1},\ldots,\mathcal{C}_{K}$. In CwC, each class group is assigned a goodness score 

\begin{equation}
    S^{(k)}_{CwC}(a) = \sum_{c_{out} \in \mathcal{C}_k} \sum_{\xi} a_{c_{out}}(\xi)^2.
    \label{eq:goodness_cwc}
\end{equation}
Each score $S^{(k)}$ evaluates local classification confidence by measuring the total activation energy within its class-specific channel group. The class $k$ yielding the highest goodness score $S^{(k)}$ therefore represents the local prediction of the layer. Because this score depends only on the total activation energy, a spatially uniform map scores as highly as a structured one, and CwC layers may develop nearly uniform activation maps that carry little of the spatial structure convolutional layers are meant to extract (Supplementary Note~\ref{supp:Noise_filtering}).

CwSG preserves the architecture of CwC but replaces raw activation energy with spatial-gradient energy. Fig.~\ref{fig:Fig1}c illustrates the distinction: CwC can reward a broad response, whereas CwSG rewards spatial variation within the feature map.
The goodness score function becomes the discrete Dirichlet energy corresponding to the $H^1$ seminorm of the activation maps

\begin{equation}
    \begin{split}
    S^{(k)}_{CwSG}(a) & = \sum_{c_{out} \in \mathcal{C}_k} \sum_{\xi} \|\nabla a_{c_{out}}(\xi)\|^2 \\
    & = \sum_{c_{out} \in \mathcal{C}_k} \sum_{h,w} \left((\partial_h a_{c_{out}})^2 + (\partial_w a_{c_{out}})^2 \right),
    \end{split}
\label{eq:goodness_cwsg}
\end{equation}
with the discrete partial derivatives defined as 

\[
\begin{cases}
    \partial_h a_{c_{out}} & = a_{c_{out}}(h+1, w) - a_{c_{out}}(h-1, w)\,, \\
    \partial_w a_{c_{out}} &= a_{c_{out}}(h, w+1) - a_{c_{out}}(h, w-1).
\end{cases}
\]

Fig.~\ref{fig:Fig1}d shows how these goodness scores provide both a prediction and a learning signal within the layer. A softmax converts the scores into class probabilities, and the layer minimizes the cross-entropy loss for the ground-truth label $y$:
\begin{equation}
\mathcal{L} = -\log \left(\mathrm{softmax}(S)_{y}\right).
\label{eq:local_ce}
\end{equation}

Differentiating the CwSG spatial objective transforms the gradient from the amplitude scaling of CwC into a discrete Laplacian (Supplementary Note~\ref{supp:Gradients}), yielding a local weight update proportional to the negated spatial second derivative of the activations

\begin{equation}
    \frac{\partial\mathcal{L}_{_{CwSG}}}{\partial W}
    = \underbrace{\delta_{CE, k}\, \big(-2\nabla^2 a\big)}_{\text{layer output}}
    \;\cdot\; \underbrace{x_{c_{in}}}_{\text{layer input}}\,,
\label{eq:final_unnorm_cwsg}
\end{equation}
where $\delta_{CE, k}$ is the cross-entropy error of the score vector, and $(\cdot)$ denotes the 2D cross-correlation. 

In our convolutional architecture, a parameter-free root-mean-square normalization (RMSnorm) is applied to the pre-activations across output channels at each spatial position, before the ReLU activation (Fig.~\ref{fig:Fig1}b). This operation fixes the channel energy at each spatial position, so a class group can raise its goodness score only by winning a larger share of a fixed budget, not by inflating amplitudes. Mathematically, the RMSnorm introduces an additional factor into the gradient:

\begin{equation}
    \frac{\partial\mathcal{L}_{_{CwSG}}}{\partial W}
    = \rho_{rms} \; \delta_{CE, k}\, \big(-2\nabla^2 a\big)
    \;\cdot\; x_{c_{in}}\,,
\label{eq:supp_final_cwsg}
\end{equation}

where $\rho_{rms}$ is a normalization matrix that enforces this cross-channel competition (Supplementary Note~\ref{supp:Gradients}). The gradient depends only on the active layer's input, forward activations, and target label: no error is transported between layers and no transposed weight read is required. Batched training retains only the active layer's local buffers, rather than activations from the full network (see Methods).

\begin{table}[h!]
\centering
\caption{\textbf{Full-precision optimization test accuracy of CNN-7 on MNIST \cite{lecun2010mnist}, Fashion-MNIST \cite{xiao2017fashion}, and CIFAR-10 \cite{Krizhevsky09cifar10}.} BP and DFA are trained end-to-end, and the layer-wise methods with the greedy schedule, under the matched per-parameter update budget defined in Methods ($N{=}50$ epochs) with stochastic gradient descent (SGD). The batch size is 64. Values are mean $\pm$ one standard deviation over five independent seeds after hyperparameter selection on the validation split (Supplementary Note~\ref{supp:HPO}). Bold marks the highest mean accuracy among backpropagation-free methods.}
\label{tab:FP32_comparison}
\setlength{\tabcolsep}{7pt}
\begin{tabular}{lccc}
\toprule
Algorithm & MNIST & Fashion-MNIST & CIFAR-10 \\
\midrule
BP           & \acc{99.51}{0.05} & \acc{93.27}{0.19} & \acc{86.28}{1.60} \\
\midrule
DFA          & \acc{89.48}{0.43} & \acc{80.50}{1.98} & \acc{38.65}{1.06} \\
DCL          & \acc{98.97}{0.07} & \acc{91.41}{0.78} & \acc{77.28}{1.27} \\
l-DCL        & \acc{98.95}{0.15} & \acc{91.06}{0.65} & \acc{79.11}{0.12} \\
CwC          & \acc{99.18}{0.06} & \acc{91.07}{0.64} & \acc{77.77}{0.67} \\
\midrule
\textbf{CwSG (This work)}  & \textbf{\acc{99.34}{0.04}} & \textbf{\acc{92.70}{0.06}} & \textbf{\acc{83.66}{1.24}} \\
\bottomrule
\end{tabular}
\end{table}

We first evaluate CwSG under ideal numerical updates with a stochastic gradient descent (SGD) optimizer, before introducing any device constraint. All the benchmarked methods use the same CNN-7 architecture (see Methods). The layer-wise methods are trained following the greedy schedule: when layer $\ell$ is under training, earlier layers are frozen, later layers are not evaluated, and only layer $\ell$ is updated (Fig.~\ref{fig:Fig1}b). This matches the per-parameter budget of the end-to-end baselines (see Methods). All accuracies in this work are averaged over five independent seeds after hyperparameter tuning (see Methods).

As shown in Table~\ref{tab:FP32_comparison}, CwSG is the strongest backpropagation-free method on every convolutional benchmark. It reaches $99.34\%$ on MNIST, $92.70\%$ on Fashion-MNIST, and $83.66\%$ on CIFAR-10. This performance trails end-to-end BP by only 0.17, 0.57, and 2.62 percentage points, respectively.

The comparison with CwC isolates the effect of changing the local goodness while preserving the same class-to-channel encoding and greedy schedule. CwSG improves over CwC by 0.16 points on MNIST, 1.63 points on Fashion-MNIST, and 5.89 points on CIFAR-10, with the accuracy advantage increasing as the task demands richer spatial representations. Spectral analysis (Supplementary Note~\ref{supp:spectral_energy_mnist_stl}) confirms the corresponding representation change: CwC activations progressively collapse toward the near-DC band with depth, whereas CwSG retains broadband spatial structure and visibly preserves edges and textures in deep activation maps. On an MLP architecture (see Methods), where neighboring units have no intrinsic geometric relationship, the one-dimensional difference objective depends on an arbitrary ordering of the hidden units and provides no systematic advantage. We therefore treat this experiment as a negative control (Supplementary Note~\ref{supp:mlp_negative_control}).


\begin{figure}[h!]
    \centering    
    \includesvg[inkscapelatex=false,width=\textwidth]{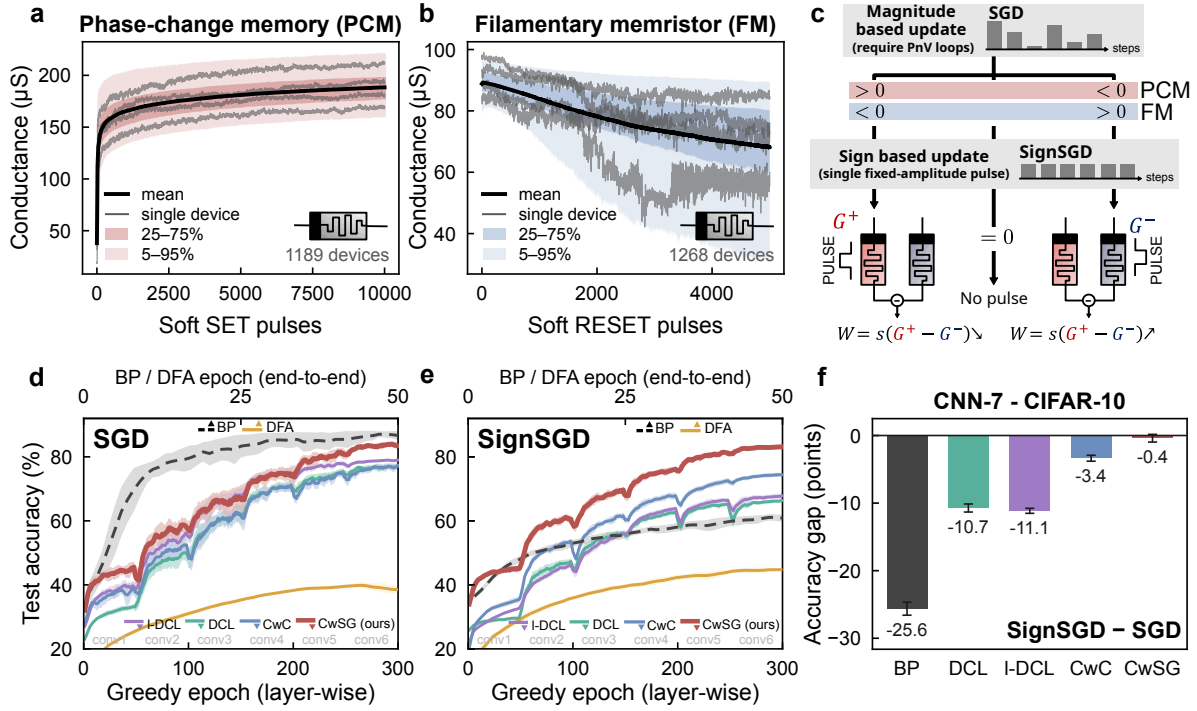}
    \caption{
    \textbf{CwSG preserves convolutional accuracy under sign-constrained updates motivated by measured analog programming.}
    \textbf{a}, \textbf{b} Measured conductance trajectories under repeated pulses for 1,189 phase-change memory devices (\textbf{a}) and 1,268 filamentary memristors (\textbf{b}). Fixed-amplitude pulses produce stochastic, state-dependent steps (see Methods, statistics detailed in Fig.~\ref{fig:Fig4}).
    \textbf{c} Optimizer mapping to a differential conductance pair. Programming pulses target $G^{+}$ or $G^{-}$ based strictly on gradient sign.
    \textbf{d} CIFAR-10 training trajectories on CNN-7 under full-precision SGD. Layer-wise algorithms follow the bottom greedy epoch axis, where ticks mark successive layers, while BP and DFA follow the top end-to-end axis. All methods train under the matched per-parameter update budget defined in Methods. Lines and shaded regions denote the mean and one standard deviation over five independent seeds.
    \textbf{e} CIFAR-10 training trajectories under SignSGD, following the layout and matched update budget of \textbf{d}.
    \textbf{f} Absolute difference in endpoint test accuracy between panels \textbf{d} and \textbf{e}. DFA is omitted because its weak full-precision baseline precludes meaningful measurement.}
    \label{fig:Fig2}
\end{figure}

\subsection*{CwSG tolerates sign-only weight updates}\label{main:Results:SignUpdates}

We next test whether CwSG can learn when weight updates no longer carry the gradient magnitude. In analog memories, precise conductance targets can be reached using program-and-verify (PnV) loops, which alternate programming pulses and readout operations until the target is reached \cite{Hermes_pcm_inf, esmanhotto2022experimental, rao2023thousands}. But this precision comes at the cost of high bit-line voltages, multiple pulse-and-read cycles per update, and reduced device endurance and retention \cite{Jiang2025Devices, renaudineau2026}.

An appealing alternative is to exploit the soft programming regime of the devices, in which each update is a single low-voltage pulse without a verification read \cite{li2018efficient, imbert2024forward, renaudineau2026}.

The measured conductance trajectories in Figs.~\ref{fig:Fig2}a and~b illustrate this regime in phase-change memory and filamentary memristors. Repeated identical pulses produce gradual conductance evolution, but the individual steps are stochastic and state-dependent, so a fixed programming pulse does not provide a precise conductance increment proportional to the gradient.

We encode each signed weight in a differential pair of devices such that $W = s(G^+-G^-)$, where $s$ is a system-level gain hyperparameter. Applying a single fixed-amplitude soft programming pulse then restricts the updates to a ternary regime: the positive device ($G^+$) is programmed, the negative device ($G^-$) is programmed, or the synapse remains unchanged (Fig.~\ref{fig:Fig2}c). This discrete operation discards the continuous magnitude of the computed gradient, leaving only its sign to determine the update.

To separate the loss of gradient magnitude from device variability, we first use sign-only stochastic gradient descent (SignSGD) \cite{bernstein2018signsgd}. For a gradient component $g_{ij}=\partial\mathcal{L}/\partial W_{ij}$, it replaces the SGD update $\Delta W_{ij}=-\eta g_{ij}$ with 
\begin{equation}
    \Delta W_{ij}=-\eta\operatorname{sign}(g_{ij}).
    \label{eq:signsgd_main}
\end{equation}
Every nonzero update now has magnitude $\eta$. Weights and gradient computations remain at full precision, and no device noise or state-dependent response is included at this stage.

\begin{table}[h!]
\centering
\caption{\textbf{Sign-only update test accuracy of CNN-7 on MNIST, Fashion-MNIST, and CIFAR-10.} The protocol is identical to that of Table~\ref{tab:FP32_comparison}, except that the optimizer is SignSGD.}
\label{tab:SignSGD_comparison}
\setlength{\tabcolsep}{7pt}
\begin{tabular}{lccc}
\toprule
Algorithm & MNIST & Fashion-MNIST & CIFAR-10 \\
\midrule
BP           & \acc{99.15}{0.14} & \acc{88.60}{1.49} & \acc{60.65}{1.43} \\
\midrule
DFA          & \acc{94.99}{0.68} & \acc{81.74}{0.90} & \acc{44.82}{0.30} \\
DCL          & \acc{97.88}{0.12} & \acc{88.87}{0.22} & \acc{66.56}{0.41} \\
l-DCL        & \acc{97.89}{0.23} & \acc{89.67}{0.22} & \acc{67.96}{0.82} \\
CwC          & \acc{98.24}{0.22} & \acc{90.17}{0.29} & \acc{74.41}{0.72} \\
\midrule
\textbf{CwSG (This work)}  & \textbf{\acc{98.71}{0.09}} & \textbf{\acc{91.22}{0.19}} & \textbf{\acc{83.26}{0.32}} \\
\bottomrule
\end{tabular}
\end{table}

Under SignSGD, CwSG reaches $98.71\%$ on MNIST, $91.22\%$ on Fashion-MNIST, and $83.26\%$ on CIFAR-10 (Table~\ref{tab:SignSGD_comparison}). Relative to its SGD results, the corresponding losses are only 0.63, 1.48, and 0.40 percentage points. The excellent robustness of CwSG against sign-only updates, compared with other methods, is clearest on CIFAR-10, where CwC loses 3.36 points, DCL and l-DCL each lose approximately 11 points, and BP loses 25.63 points (Figs.~\ref{fig:Fig2}d--f). Changing the local
objective therefore improves not only the accuracy reached under SGD, but also how much of that accuracy is retained when update magnitudes are removed.

Two features help interpret these results; class-to-channel supervision and the spatial form of the local objective. The first feature is shared by CwC and CwSG. For a single example without RMSnorm, the weight gradient factorizes into 
\begin{equation}
    \frac{\partial \mathcal{L}}{\partial W} = \underbrace{\delta_{CE,k}}_{\substack{\text{supervised} \\ \text{error}}}\,\underbrace{\varphi}_{\substack{\text{Hebbian} \\ \text{factor}}},
    \label{eq:gradient_factorization}
\end{equation}
where  $\varphi$ is a Hebbian factor computed from the forward pass. The cross-entropy factor $\delta_{CE,k}$ is negative for the target class and positive for all other classes. Consequently, the label determines whether SignSGD follows or opposes the sign of the local factor $\varphi$. Each signed update therefore still reinforces the target group and weakens the others; only the magnitude information, which expresses confidence rather than direction, is discarded.

Despite sharing this sign-sufficiency mechanism, CwC and CwSG differ in how their Hebbian factors distribute update directions across the weights. For  ReLU (non-negative) layer inputs and activations, CwC's factor $\varphi=2a\cdot x_{c_{in}}$ is non-negative (Supplementary Note~\ref{supp:Gradients}). Its nonzero weight updates therefore all point upward for a target group and downward for a non-target group. 
On a balanced $K$-class dataset, each group encounters $K-1$ times as many non-target as target examples: nine to one for ten classes. SGD compensates by giving the rarer upward updates larger magnitudes than the more frequent downward updates, but SignSGD removes this weighting, and leaves CwC with a net downward drift.

CwSG instead asks the layer to strengthen or weaken spatial contrasts, rather than simply raise or lower its overall response. Adding a uniform offset to an activation map leaves its spatial differences, and therefore its goodness, unchanged. Accordingly, the goodness derivative sums to zero over the feature map:
\begin{equation}
    \sum_{\xi}
    \frac{\partial S^{(k)}_{CwSG}}{\partial a_{c_{out}}(\xi)} =
    -2\sum_{\xi}(\nabla^2a_{c_{out}})(\xi) =0 \,,
    \label{eq:zero_mean_activation_gradient}
\end{equation}
with the finite-difference boundary convention defined in Supplementary Note~\ref{supp:Gradients}.

To isolate the consequence for weight updates, consider the Hebbian factor before ReLU masking,
\begin{equation}
    \begin{split}
    \varphi
    &= \big(-2\nabla^2a\big)\cdot x_{c_{in}}\\
    &= \underbrace{
        \bar{x}\sum_{\xi}\big(-2\nabla^2a\big)(\xi)
       }_{=\,0}
       +\big(-2\nabla^2a\big)\cdot
        \big(x_{c_{in}}-\bar{x}\big)\, ,
    \end{split}
    \label{eq:hebbian_split}
\end{equation}
where the correlation is evaluated at a fixed kernel offset, and $\bar{x}$ is the mean of the corresponding input values. The mean term vanishes by Eq.~\ref{eq:zero_mean_activation_gradient}, leaving a contribution determined by how input variations align with the activation contrasts. Unlike the non-negative CwC factor, this contribution can take either sign.
A non-target example therefore no longer forces a group-wide decrease of weights, and sign-only updates can retain direction to shape the response, instead of only turning its overall activity up or down (Supplementary Note~\ref{supp:SignUpdateProperty}).

This argument identifies a source of robustness and not a formal guarantee of drift-free weight updates. ReLU masking, RMSnorm, and batch averaging modify the sign structure; all reported simulations include these operations and use the full gradients (Supplementary Note~\ref{supp:Gradients}).

BP layers lack this structural mechanism because the error is a weighted mixture of downstream class errors. Under SignSGD, fixed-amplitude updates in upper layers can flip the signs of downstream gradients, causing a compounding failure with depth.

\begin{figure}[h!]
\centering
\includesvg[inkscapelatex=false, width=\textwidth]{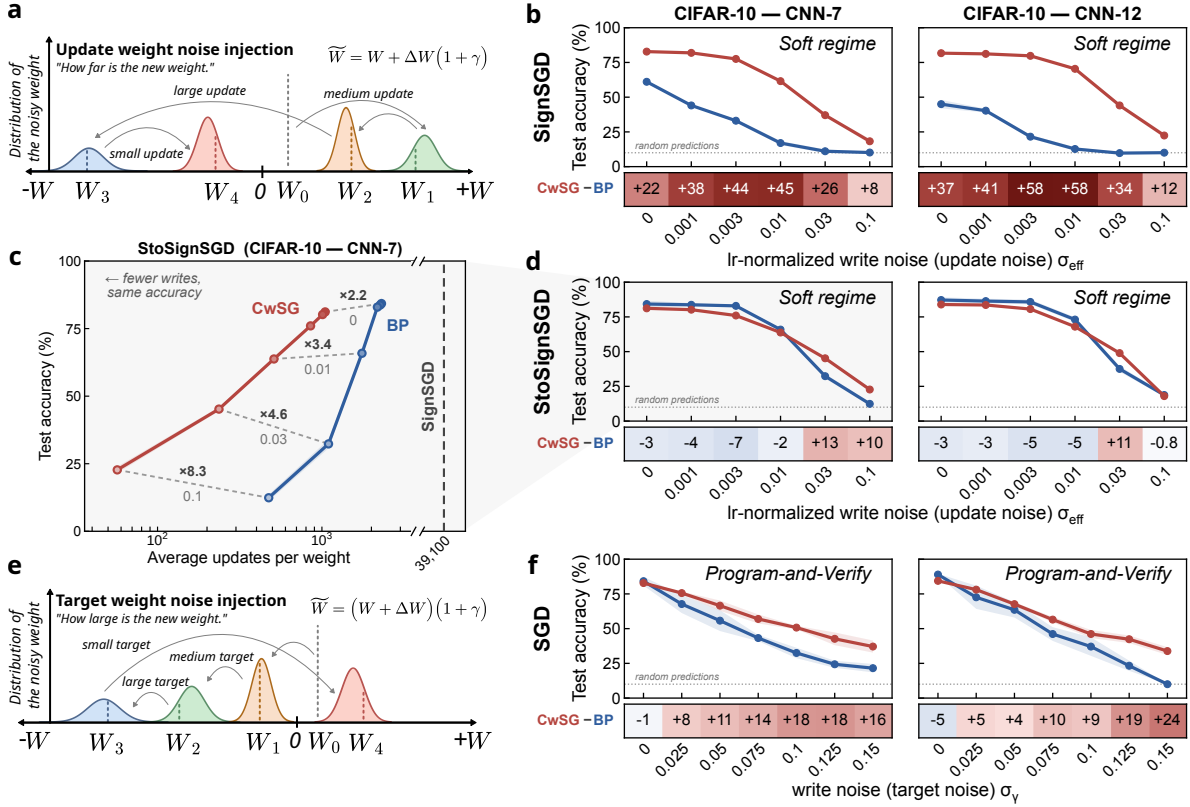}
\caption{
    \textbf{CwSG requires fewer weight updates than backpropagation (BP) and is intrinsically more robust to write noise.} 
    \textbf{a} Update-weight noise injection model with $\gamma \sim \mathcal{N}(0, \sigma_\gamma^2)$ emulating state-blind soft SET and RESET programming. 
    \textbf{b} Test accuracy against learning-rate-normalized update noise $\sigma_{\mathrm{eff}} = \eta\sigma_\gamma$ for CNN-7 and CNN-12 on CIFAR-10 under SignSGD. The curves show the average test accuracy over five independent runs, and the panels below display the absolute accuracy difference between CwSG and BP. Hyperparameters are optimized for the noiseless condition of each rule and kept identical under noise. 
    \textbf{c} Test accuracy against the average number of updates per weight for the CNN-7 StoSignSGD sweep shown in \textbf{c}. Dashed guides connect matched physical noise levels and are annotated with the update ratio between BP and CwSG. 
    \textbf{d} Test accuracy against learning-rate-normalized update noise under StoSignSGD, following the conventions of \textbf{b}. 
    \textbf{e} Target-weight noise injection model emulating iterative program-and-verify schemes. 
    \textbf{f} Test accuracy under full-precision SGD against target noise $\sigma_\gamma$, following the conventions of \textbf{b}.}
\label{fig:Fig3}
\end{figure}

\subsection*{CwSG limits sensitivity to write noise}\label{main:Results:NoiseRobustness}

Retaining accuracy under fixed-magnitude updates does not yet establish robustness to errors in those updates. We next compare BP and CwSG on CIFAR-10 using CNN-7 and CNN-12 under controlled Gaussian programming-noise models \cite{Joshi2020,Rasch21}, pairing each optimizer with the noise of the programming scheme it represents (see Methods).

For SignSGD, we add noise to each applied weight increment: $\widetilde{W} = W + \Delta W(1 + \gamma)$ with $\gamma \sim \mathcal{N}(0,\sigma_{\gamma}^2)$. For a nominal magnitude $\eta$, the weight standard deviation per committed write is $\sigma_{\mathrm{eff}} = \eta\sigma_\gamma$. We therefore compare the learning rules at the same $\sigma_{\mathrm{eff}}$ (see Methods).

Fig.~\ref{fig:Fig3}b reports the results of this first study with hyperparameters selected as detailed in Supplementary Note~\ref{supp:HPO}. Under SignSGD, as discussed in the previous section, BP is penalized before any noise is injected: its noise-free accuracy falls to $60.6\%$ on CNN-7 and $45.0\%$ on CNN-12, down from $86.3\%$ and $88.9\%$ under full-precision SGD. As the noise variance increases, the compounding error overwhelms the global gradient, dropping BP to near random chance by $\sigma_{\mathrm{eff}} = 0.03$ on CNN-7 and by $0.01$ on CNN-12. CwSG retains substantially more accuracy. At $\sigma_{\mathrm{eff}} = 0.01$, it maintains $61.5\%$ on CNN-7 and $70.4\%$ on CNN-12, and it approaches chance level only at $\sigma_{\mathrm{eff}} = 0.1$.

SignSGD commits a pulse for every candidate update. To reduce this programming cost, we next test a stochastically gated SignSGD (StoSignSGD) optimizer, which issues each pulse with a probability that increases with the gradient magnitude:
\begin{equation}
\Delta W_{ij} = -\eta\operatorname{sign}(g_{ij})b_{ij}\,, \quad b_{ij} \sim \operatorname{Bernoulli} \left[ \operatorname{clip} \left( \theta|g_{ij}|,0,1 \right) \right],
\label{eq:stosignsgd_main}
\end{equation}
where $\theta$ is a gating gain hyperparameter governing saturation. Every committed write remains a fixed-magnitude event, while the expected update in the unsaturated regime is directly proportional to the continuous gradient (see Methods). In the literature, stochastic pulse-coding schemes exploit the same probabilistic quantization but encode the update magnitude in the number of applied pulses \cite{Gokmen17rpu}, whereas here it is carried entirely by the probability of committing a single pulse or none.

The results using this StoSignSGD (Fig.~\ref{fig:Fig3}d) substantially change this picture with regard to SignSGD. Because StoSignSGD recovers the continuous gradient in expectation, it restores BP to near its noise-free baseline ($84.3\%$ on CNN-7, $87.2\%$ on CNN-12) and gives it a lead over CwSG of up to 6.9 percentage points in the low-noise regime ($\sigma_{\mathrm{eff}} \leq 0.01$). This advantage does not survive high programming noise, and the hierarchy inverts at $\sigma_{\mathrm{eff}} = 0.03$ on both architectures, where CwSG establishes a lead of more than 10 percentage points on CNN-7 and maintains it through the end of the sweep. The gating also benefits CwSG itself, improving its accuracy at $\sigma_{\mathrm{eff}} = 0.03$ on CNN-7 from $37.0\%$ under SignSGD to $45.2\%$, as committing fewer updates reduces the accumulated write noise for both learning rules.

StoSignSGD also shows a strong programming-cost advantage (Fig.~\ref{fig:Fig3}c). Compared to SignSGD, which applies 39,100 updates per weight, StoSignSGD reduces the total number of applied pulses by a factor of 17 under BP (2,303 updates per weight). CwSG sparsifies the updates even further, achieving a 37.4-fold reduction (1,046 updates per weight). At matched noise levels, CwSG therefore reaches comparable accuracy while committing roughly two to eight times fewer pulses per weight, a reduction that translates directly into programming-energy savings.

We next test whether the noise-resilience advantage of CwSG persists when updates retain their magnitudes directly through SGD. For this study, we use target noise, which perturbs the desired stored weight rather than the increment (Fig.~\ref{fig:Fig3}e):  $\widetilde{W} = (W + \Delta W)(1 + \gamma)$. This equation provides a simplified model of residual error after programming a conductance target, as in PnV-based schemes. Here, $\sigma_\gamma$ measures fractional error relative to the target weight. Fig.~\ref{fig:Fig3}f shows that under SGD, BP and CwSG start on par in the noise-free case, but CwSG overtakes BP for $\sigma_\gamma \geq 0.025$, and its lead widens substantially with the noise level. At $\sigma_\gamma = 0.15$, BP falls to chance level on CNN-12, while CwSG still reaches $33.9\%$. 

The same hierarchy extends to readout noise ($\sigma_{\mathrm{rd}} = 0.1$) for all three optimizers; on CNN-7, CwSG maintains up to a 45-point advantage over BP across the entire SignSGD update-noise grid (Supplementary Note~\ref{supp:read_noise}).

This hierarchy stems from the dataflow of BP, where gradients are computed through cascading transposed weight matrices, so every write perturbation persists as an additive error that re-enters subsequent backward passes and compounds geometrically with depth (Supplementary Note~\ref{supp:Noise_prop}). The local updates of CwSG decouple the layers and bypass this inter-layer noise propagation entirely. In addition, the discrete Dirichlet energy acts as a spatial high-pass filter that suppresses the low-frequency spatial noise imprinted by imperfect updates, yielding a higher signal-to-noise ratio than the raw amplitude goodness of CwC (Supplementary Note~\ref{supp:Noise_filtering}).

\subsection*{CwSG trains with measured PCM and FM programming trajectories}\label{main:Results:Devices}

\begin{figure}[h!]
\centering
\includesvg[inkscapelatex=false, width=\textwidth]{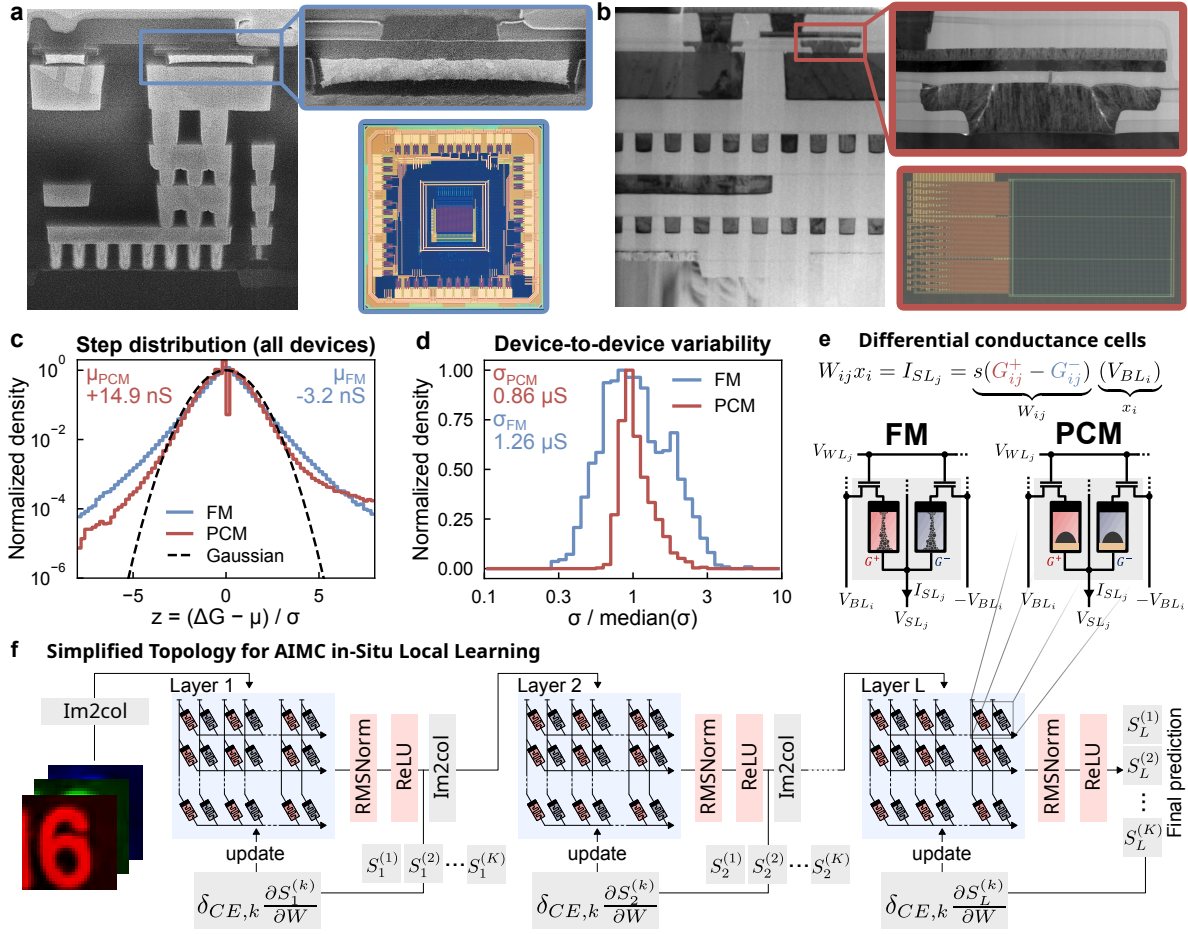}
\caption{
    \textbf{Physical substrate and statistical characterization of fabricated memory arrays.}
    \textbf{a}, \textbf{b} Transmission electron microscopy images of a filamentary memristor from the characterized array (\textbf{a}) and of a representative phase-change memory device of the same technology generation (\textbf{b}), both integrated in the back end of line of a hybrid memristor--CMOS process, alongside optical micrographs of their respective 1T1R arrays (see Methods).
    \textbf{c} Distribution of normalized conductance steps $z$ across all devices.
    \textbf{d} Device-to-device variability of the per-device step standard deviation normalized to the median.
    \textbf{e} Differential-pair weight mapping $W = s(G^+ - G^-)$, where $s$ is a system-level gain hyperparameter. The illustrated electrical connections route a single, fixed-amplitude, sub-1~V programming pulse (see Methods) to either $G^+$ or $G^-$.
    \textbf{f} Crossbar implementation for CwSG training. Each convolutional layer unrolled via im2col maps onto its own memristive array. The RMSnorm and ReLU activation are computed digitally, allowing each layer to produce its per-class goodness $S_l^{(k)}$ locally for both weight updates and final prediction.}
\label{fig:Fig4}
\end{figure}

We next test whether the advantages observed under Gaussian noise persist when synaptic updates follow measured device responses. The simulations use programming trajectories recorded from 1,189 fabricated phase-change memory (PCM) devices and 1,268 filamentary memristors (FM) (Fig.~\ref{fig:Fig4}a and~b; see Methods).

For learning, each technology is operated in a single programming direction. In PCM, successive low-voltage soft SET pulses increase conductance through progressive crystallization, with a steep initial rise followed by saturation near $190~\mu$S over the measured 10,000-pulse sequence (Fig.~\ref{fig:Fig2}a). In FM, soft RESET pulses progressively decrease conductance through partial dissolution of the conductive filament. The measured 5,000-pulse trajectories begin near $90~\mu$S and show a more linear decrease (Fig.~\ref{fig:Fig2}b).

The mean single-pulse conductance changes are $+14.9$~nS for PCM and $-3.2$~nS for FM. Both technologies exhibit heavy-tailed step distributions, with large increments occurring substantially more often than in the Gaussian reference (Fig.~\ref{fig:Fig4}c). The spread of the increments also varies markedly between devices (Fig.~\ref{fig:Fig4}d). These measurements therefore introduce state-dependent programming behavior, heavy-tailed programming steps and device-specific variability into the learning evaluation.

Each simulated weight is represented by a differential pair, $W=s(G^+-G^-)$, using trajectories from two distinct measured devices (Fig.~\ref{fig:Fig4}e). Because the networks contain many more weights than the characterized device pools, measured trajectories are reused approximately equally across the network, but a trajectory is never paired with itself (see Methods). Fig.~\ref{fig:Fig4}f illustrates the corresponding crossbar mapping: each convolutional layer maps to a dedicated array region. 

We train the wide CNN-7W architecture (see Methods) on MNIST and SVHN with StoSignSGD. Hyperparameters are optimized independently per algorithm and per technology. The update gating gain $\theta$ controls the probability of applying a programming pulse.

\begin{figure}[h!]
\centering
\includesvg[inkscapelatex=false, width=\textwidth]{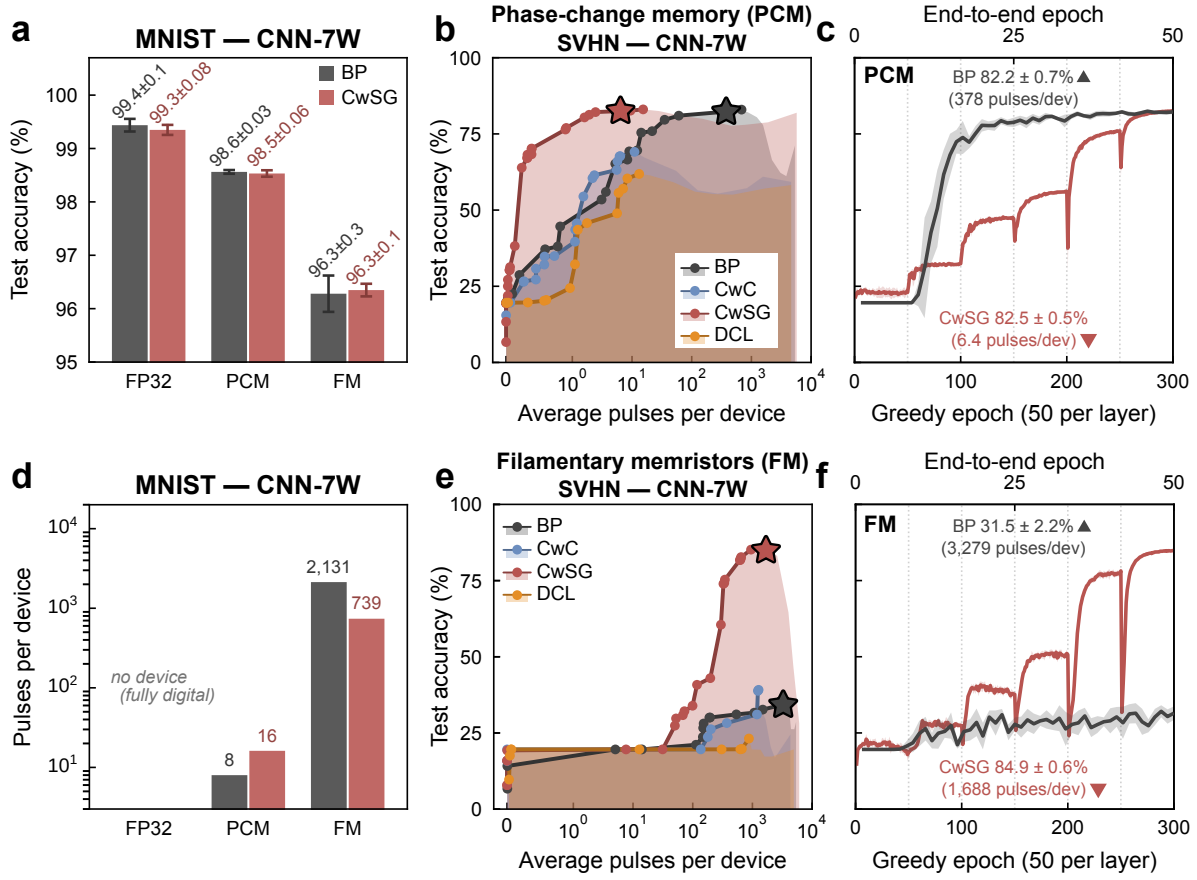}
\caption{
    \textbf{Device-calibrated training of CwSG and backpropagation (BP) on analog devices.} All panels show the CNN-7W architecture trained at batch size 256 with StoSignSGD under the matched per-parameter update budget defined in Methods: $N$ greedy epochs per layer for CwSG, CwC, and DCL and $N$ end-to-end epochs for BP, with $N{=}30$ on MNIST and $N{=}50$ on SVHN.
    \textbf{a} MNIST test accuracy.
    \textbf{b} SVHN test accuracy against average pulses per device on phase-change memory. Filled regions mark the best accuracy at each pulse budget, demonstrating that excess pulses do not yield further improvements.
    \textbf{c} SVHN test accuracy throughout training for the selected configurations marked by stars in \textbf{b} spanning 300 cumulative greedy epochs for CwSG and 50 end-to-end epochs for BP. Dips mark layer transitions (see Methods).
    \textbf{d} MNIST programming pulses per device.
    \textbf{e}, \textbf{f} SVHN test accuracy and training trajectories on filamentary memristors, following the exact layout conventions of \textbf{b} and \textbf{c}.}
\label{fig:Fig5}
\end{figure}

On MNIST (Figs.~\ref{fig:Fig5}a and~d), CwSG and BP reach similar accuracies on both technologies. The difference emerges in programming cost. On PCM, BP uses an average of 8.0 pulses per device over training, compared with 16.2 for CwSG, because the global gradients of an easily fitted task vanish after the first few epochs. On the noisier filamentary memristors, this hierarchy inverts, with CwSG requiring nearly three times fewer pulses than BP.

On the harder SVHN dataset (Fig.~\ref{fig:Fig5}b, and~c), with PCM, CwSG reaches $82.5\%$ accuracy using only 6.4 pulses per device, whereas BP reaches $82.2\%$ using 378 pulses. Both remain below their full-precision baselines of $93.6\%$ and $95.1\%$, respectively, but CwSG matches BP's accuracy with 59-fold fewer programming events. 

On FM, the benefits of CwSG extend to accuracy (Fig.~\ref{fig:Fig5}e, and~f). CwSG reaches $84.9\%$, whereas BP reaches only $31.5\%$, a gap of 53.4 percentage points. CwSG still uses fewer pulses: 1,688 per device, compared with 3,279 for BP. 

Several lessons emerge from these results. In all cases except MNIST on PCM, CwSG uses fewer programming pulses while achieving similar or higher accuracy.

On SVHN, this advantage takes two distinct forms: far fewer writes at comparable accuracy on PCM, and substantially higher accuracy with fewer writes on FM. These observations are consistent with the reduced sensitivity of CwSG to imperfect programming discussed above. By contrast, the noise-sensitivity of BP is amplified by realistic statistics: the non-ergodic, high-kurtosis device noise and asymmetric updates trap BP in a compensatory cycle, forcing it to spend additional pulses to correct the perturbations injected by its own preceding updates. Note that the accuracies reported are lower than those of the ideal model at matched noise level, indicating that realistic device statistics penalize global error transport more severely than idealized assumptions suggest.

The device comparison reveals a second lesson: the response that gives the highest accuracy depends on the learning rule. BP performs much better on PCM than on FM, whereas CwSG reaches 2.4 percentage points higher accuracy on FM. The contrasting conductance trajectories offer a possible explanation. PCM's saturating response provides progressively smaller increments, while FM maintains a more uniform update granularity. CwSG's robustness may allow it to exploit the latter despite substantial programming variability. 

The comparisons with CwC and DCL show that layer-wise training alone does not reproduce CwSG's performance: both remain below CwSG in the PCM and FM sweeps (Fig.~\ref{fig:Fig5}b, and~e). CwC is a particularly informative comparison because it shares CwSG's class-to-channel supervision and greedy schedule. The improvement therefore highlights the importance of the local objective, not simply the removal of network-wide backpropagation.

The accuracy advantage of CwSG extends to CIFAR-10 (Supplementary Note~\ref{supp:cifar10_dev_bs256}). On PCM, CwSG reaches $58.6\%$ versus $50.0\%$ for BP while using 5.1 rather than 62 pulses per device, approximately a 12-fold reduction. On FM, CwSG reaches $59.4\%$ with 1,415 pulses per device versus $40.2\%$ on BP. Thus, CwSG improves accuracy on both technologies, while the programming-cost advantage depends on the task and device response.

\subsection*{CwSG avoids quantized transposed array reads}\label{main:Results:archi}

\begin{figure}[h!t]
\centering
\includesvg[inkscapelatex=false,width=\textwidth]{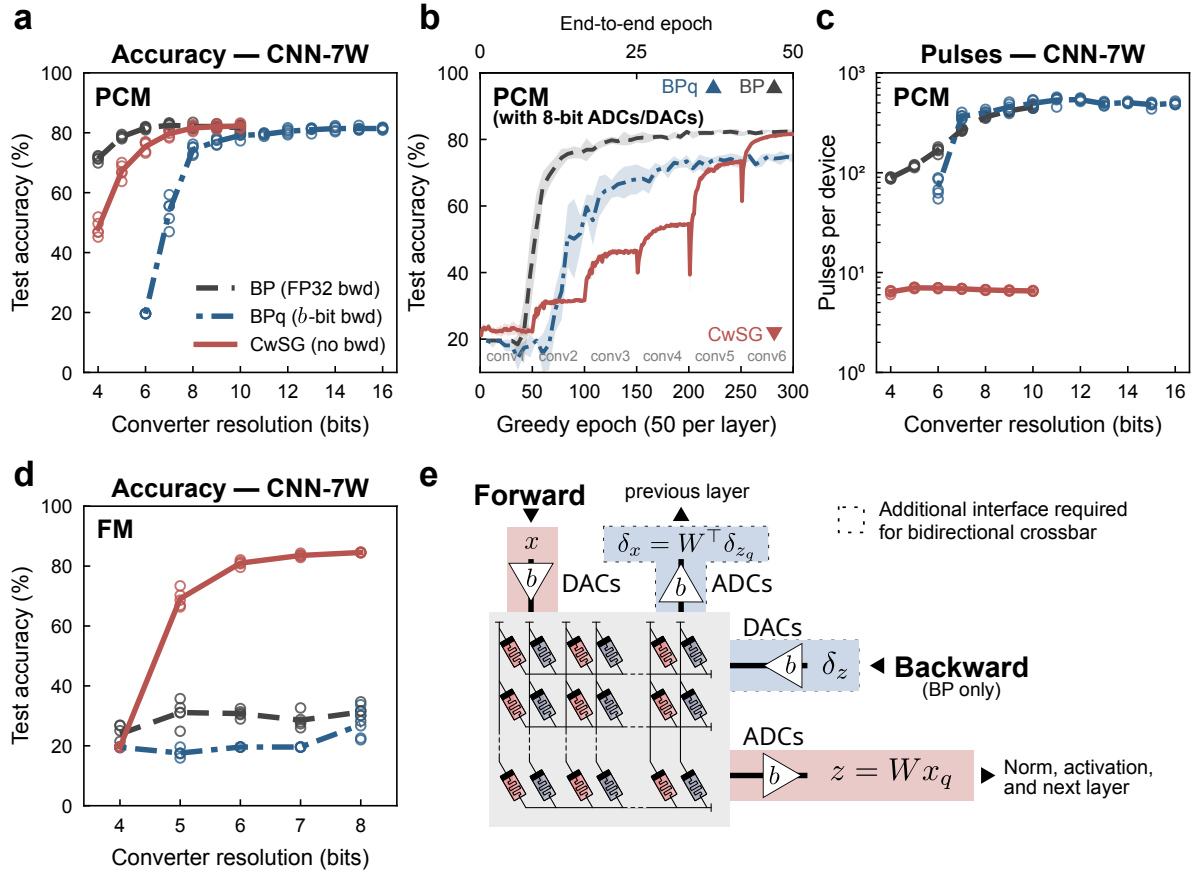}
\caption{
    \textbf{Converter resolution and the architectural cost of a quantized backward pass.}
    \textbf{a} SVHN test accuracy against converter resolution $b$ on phase-change memory. Evaluation uses CNN-7W with device-calibrated synapses, batch size 256, StoSignSGD, and 50 epochs per layer for CwSG, and 50 epochs end-to-end for BP and BPq. Plotted values show the five independent seeds and their mean. BP (FP32 bwd) and CwSG (no bwd) quantize the forward drive and readout and do not quantize any term in the gradient computation. Fully quantized BP (BPq) restricts both backward drive and readout to $b$ bits. 
    \textbf{b} Test accuracy throughout training at 8 bits, following the trajectory conventions of Fig.~\ref{fig:Fig5}c, with the layer under training labeled.
    \textbf{c} Programming pulses per device against converter resolution on phase-change memory.
    \textbf{d} SVHN test accuracy against converter resolution on filamentary memristors.
    \textbf{e} Peripheral circuitry requirements for intermediate layers. The forward pass (CwSG) requires one digital-to-analog converter per driven row and one analog-to-digital converter per column. Fully quantized BP (BPq) additionally drives the error onto the columns and reads the rows, forcing the duplication of both banks on opposite edges to create a bidirectional crossbar.}
\label{fig:Fig6}
\end{figure}

Analog training must accommodate not only imperfect programming, but also finite-resolution peripheral converters. Digital-to-analog converters (DACs) drive the array inputs, and analog-to-digital converters (ADCs) digitize the resulting outputs; their precision contributes to the accelerator's energy and area requirements \cite{wan2022compute, ambrogio2023analog}. We therefore evaluate converter quantization at the SVHN operating points of Figs.~\ref{fig:Fig5}c and~f, using the same device-calibrated synapses.

At each layer, the forward DACs and ADCs are quantized to $b$ bits, using the drive and readout ranges specified in Methods. CwSG requires only these forward array interfaces. BP additionally evaluates $W^{\top}\delta$, which requires a second bank of DACs and ADCs for driving an error signal into the array and digitizing the transposed-product output (Fig.~\ref{fig:Fig6}e).

To numerically isolate the effect of these additional conversions, we compare two BP configurations. BP (FP32 bwd) uses $b$-bit forward interfaces but leaves the backward products free of converter quantization. BPq additionally quantizes the backward drive and readout to $b$ bits. 

On PCM, eight-bit forward conversion has little effect on accuracy: CwSG and BP (FP32 bwd) remain within one percentage point of their respective ideal-converter results (Fig.~\ref{fig:Fig6}a). Quantizing BP's backward interfaces introduces an additional penalty. BPq falls to near-chance accuracy at six bits and, at eight bits, remains approximately eight percentage points below the ideal-backward BP control. The gap narrows as converter resolution increases. Thus, a forward resolution sufficient to preserve classification accuracy does not necessarily suffice for backward error transport under the same converter model. 

The eight-bit training trajectories illustrate this distinction (Fig.~\ref{fig:Fig6}b), while the pulse counts show that CwSG retains its low programming activity (Fig.~\ref{fig:Fig6}c). The accuracy advantage of CwSG on FM also persists with eight-bit forward conversion (Fig.~\ref{fig:Fig6}d). Converter quantization therefore does not erase the advantages observed with the measured device trajectories.

\subsection*{Local updates on small batches}\label{main:Results:minibatch}

Layer-local training removes the need to retain activations across the network, but a buffer still needs to memorize the activation of the active layer over a batch (see Methods). Reducing the batch size lowers this storage requirement while averaging each learning decision over fewer examples. We therefore test whether CwSG remains effective when updates are computed from small batches.

We first train the CNN-7W architecture under StoSignSGD on MNIST at batch sizes ranging from 256 down to a single sample (see Methods). Under full-precision weights, reducing the batch size barely affects either rule, with the accuracy staying within 0.7 points for CwSG (from 99.3\% to 98.6\%) and 0.4 points for BP (from 99.4\% to 99.0\%).

\begin{figure}[h!t]
\centering
\includesvg[inkscapelatex=false,width=\textwidth]{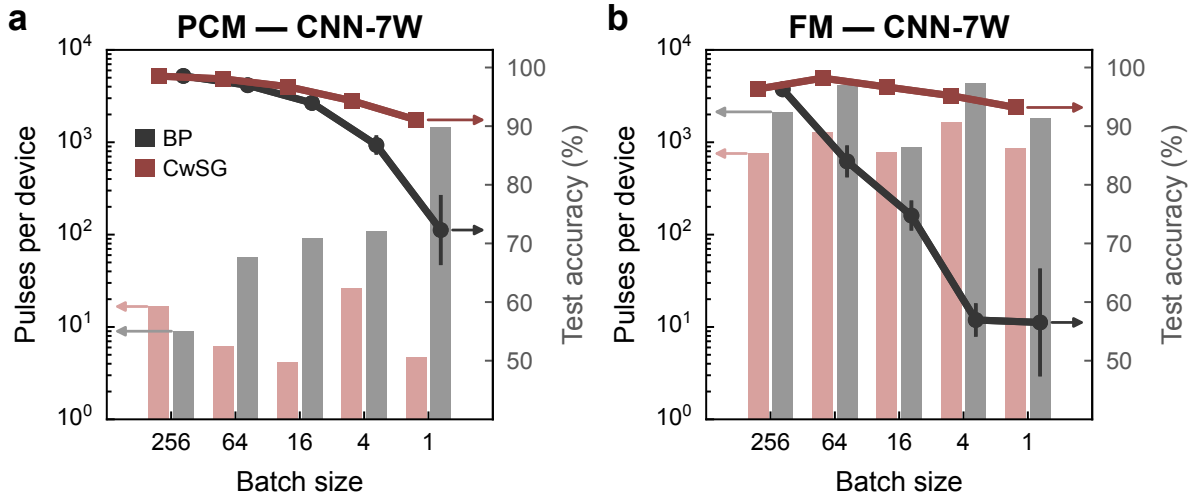}
\caption{
    \textbf{MNIST accuracy at different batch sizes for CwSG and backpropagation (BP) on analog devices.}
    \textbf{a}, \textbf{b} MNIST test accuracy and programming pulses on CNN-7W with phase-change memory (\textbf{a}), and filamentary memristors (\textbf{b}) for five independent seeds with one standard deviation represented as a vertical bar. Training is done under StoSignSGD under the matched per-parameter update budget defined in Methods: $N$ greedy epochs per layer for CwSG and $N$ end-to-end epochs for BP, with batch size 1 ($N{=}1$), 4 ($N{=}3$), 16 ($N{=}5$), 64 ($N{=}10$), and 256 ($N{=}30$).
    }
\label{fig:Fig7}
\end{figure}

The PCM simulations reveal a much larger difference (Fig.~\ref{fig:Fig7}a). BP and CwSG reach similar accuracies at batch size 256, but CwSG performs better at batch sizes 64, 16, and 4 while using between 4-fold and 29-fold fewer programming pulses. At batch size one, CwSG still reaches $91.1\%$, compared with $72.3\%$ for BP.

BP is even more sensitive in the FM simulations (Fig.~\ref{fig:Fig7}b). Both rules reach $96.3\%$ at batch size 256, but at batch size one BP falls to $56.5\%$, with a large seed-to-seed spread. CwSG retains $93.2\%$, only 3.1 percentage points below its large-batch accuracy. Thus, the MNIST single-sample configurations separate the learning rules much more strongly with measured programming trajectories than with ideal numerical updates.

The harder SVHN task reveals both the advantage and the limits of small-batch local learning (Supplementary Note~\ref{supp:batch_svhn}). With ideal numerical updates, both rules remain within 3.5 percentage points of their batch-size-256 accuracies under the tested schedules. On PCM, CwSG reaches $41.1\%$ at batch size one using approximately nine pulses per device, whereas BP falls to near-chance accuracy. Although this remains a substantial advantage, CwSG also loses considerable accuracy relative to its large-batch result. On FM, CwSG still learns at batch size 64, but both methods approach chance-level performance at batch size one.

The combination of two constraints is responsible for the collapse of BP on both technologies. The few-sample updates and stochastic programming can be accommodated individually, but together they remove the gradient averaging that shields the optimization from the update stochasticity. In these conditions, the weights undergo a continuous random walk, and BP re-amplifies the accumulated perturbation through the error propagation through the transposed weight at every gradient computation. Under the same constraints, CwSG confines the perturbations to its active layer, avoiding the propagation of the gradient (Supplementary Note~\ref{supp:Noise_prop}).

\section*{Discussion}\label{sec:Discussion}

Robust analog learning depends not only on where the learning signal is computed, but also on what each layer is trained to reward. CwSG retains the class-to-channel supervision and layer-local training of CwC while replacing activation-amplitude goodness with spatial-gradient energy. Under the common architectures and training protocols evaluated here, this change improves both convolutional accuracy and robustness to imprecise updates. On CIFAR-10, discarding gradient magnitudes reduces accuracy by only 0.40 percentage points for CwSG, compared with 3.36 points for CwC and approximately 11 points for DCL and l-DCL. The comparisons with CwC and DCL under measured device trajectories reinforce the same conclusion: layer-wise training alone does not reproduce the robustness obtained by redesigning the local objective.

These results complement our previous experimental demonstration of  supervised Forward-Forward and competitive-forward learning with single-pulse, sub-1 V RESET programming in FMs \cite{renaudineau2026}. That demonstration trained fully connected classifier layers on features from a fixed, pretrained convolutional backbone. CwSG extends learning to the convolutional representations themselves.

The spatial objective connects representation advantage quality to update structure. Amplitude-based goodness can reward a spatially uniform class-specific response, whereas adding a uniform offset leaves CwSG's goodness unchanged. Spectral analysis shows that CwSG preserves broadband spatial structure across depth (Supplementary Note~\ref{supp:spectral_energy_mnist_stl}). This offset invariance gives the goodness derivative a zero spatial sum. A simplified analysis shows how this property allows update directions to reflect spatial contrasts (Supplementary Note~\ref{supp:SignUpdateProperty}).

The absence of inter-layer backward transport provides a separate advantage, shared with other layer-local methods. It removes the dependence of an earlier layer's learning signal on downstream weights, without preventing imperfect upstream weights from altering the forward representations. CwSG therefore avoids transposed crossbar reads and their associated backward converter requirements. The converter experiments show that its accuracy can be retained with eight-bit forward interfaces. Its greedy schedule also replaces network-wide activation retention with buffers for the active layer (see Methods), creating a memory saving that could be used either to reduce on-chip storage or to increase batch averaging.

The comparison between PCM and FM reveals a distinct accuracy-programming-cost trade-off: PCM requires fewer learning pulses on all three datasets and also provides higher accuracy on MNIST. FM increases mean accuracy by 2.4 percentage points on SVHN and 0.8 points on CIFAR-10, but requires approximately 264 and 277 times as many pulses, respectively. Among these configurations, PCM is the better choice when minimizing programming activity, whereas FM provides a modest additional accuracy margin on the two harder tasks.

In the literature, analog training schemes using high-precision digital accumulators \cite{LeGallo2017,Nandakumar2020} or auxiliary analog arrays \cite{Gokmen20_tikitaka,Rasch2024} compensate for stochastic programming while retaining end-to-end backpropagation. CwSG supports learning through direct, unverified programming on forward crossbars, without auxiliary arrays or program-and-verify circuitry. Nevertheless, compensation mechanisms could also be combined with local learning to improve update fidelity.

Several algorithmic constraints delimit the present results. The demonstrated advantage relies on spatially structured feature maps; the one-dimensional analogue tested on fully connected networks provides no systematic improvement (Supplementary Note~\ref{supp:mlp_negative_control}). The class-to-channel encoding also requires at least one channel per class in every supervised layer, making narrow layers difficult to reconcile with large class counts. Class grouping \cite{Papachrist24ClusterF} and adaptive assignments \cite{ghader2026forwardonly, Papachristodoulou2026ChaCo} provide possible directions for relaxing this constraint, but their compatibility with stochastic programming remains to be evaluated.

Local gradient formation, normalization, goodness evaluation, and stochastic pulse selection remain peripheral operations, evaluated in full precision here (Supplementary Note~\ref{supp:Gradients}). Their arithmetic, memory traffic, and random-number generation, together with the additional forward passes through the frozen layers that the greedy schedule requires, must be included when translating reduced pulse counts into system-level energy savings. Intrinsically stochastic magnetic tunnel junctions could be explored as a source for the stochastic pulse selection of StoSignSGD \cite{Iftakher2026}. Our evaluation replays measured single-device trajectories. Closed-loop training on an integrated crossbar, with array-level parasitics and the local learning periphery in place, is the natural next step.

Taken together, these results show that precise programming is not the only route to effective analog learning. CwSG illustrates how matching each layer's objective to spatial representations and simple programming operations supports learning with imperfect analog synapses.

\section*{Acknowledgements}

This work benefited from France 2030 government grants managed by the French National Research Agency (ANR-22-PEEL-0010, ANR-22-PEEL-0013, ANR-23-PEIA-0002). 
This project was provided with computing HPC and storage resources by GENCI at IDRIS thanks to the grant 2025-AD011016471 on the supercomputer Jean Zay's A100-H100 partition. Large language models (OpenAI ChatGPT, Anthropic Claude) were used for copyediting parts of this manuscript.

\section*{Author contributions statement}
T.D. proposed the spatial gradient objective, developed the software framework, and performed the full-precision and device-calibrated simulations.
A.R. conducted the analysis and characterization of the filamentary memristor soft RESET regime. 
G.N., G.B., and J.M.L. fabricated the phase-change memory devices and conducted the soft SET programming regime analysis. 
A.R. and J.M.L. performed the phase-change memory characterization. 
J.M.P., J.O.K., and D.Q. supervised the project. 
T.D. wrote the initial version of the manuscript. 
All authors discussed the results and contributed to the final manuscript.

\section*{Competing interests}
Damien Querlioz is a Guest Editor of the Collection Future Computing: Paradigms and Design for Next Generation Applications. Damien Querlioz was not involved in the journal’s review of, or decisions related to, this paper. The remaining authors declare no competing interests.

\section*{Data availability}
The data measured in this study are available from the corresponding author upon request.

\section*{Code availability} 
The software programs developed for this work are available online: \url{https://github.com/INTEGNANO/LL-CwSG}.

\section*{Methods}\label{methods}
\subsection*{Network architectures}

\begin{table}[h!]
\centering
\caption{\textbf{Architecture conventions.} All variants share the same feature block (bias-free $3\times3$ convolution or dense projection, parameter-free RMS normalization, ReLU). Convolutional variants downsample with $2\times2$ max pooling. The parameter count covers the bias-free feature block (weights of the convolutional/hidden layers). For BP and DFA, the linear readout adds a negligible, dataset-dependent $C_{\text{last}}\times n_{\text{classes}}$ term (e.g.\ $2{,}560$ for CNN-7 at 10 classes). CNN first-layer parameter counts assume a 3-channel input, while MLP-5 assumes a flattened $28\times28$ input.}
\label{tab:archs}
\begin{tabular}{lcccl}
\toprule
Architecture & Hidden layers & Pooling & Parameters & Hidden layer width\\
\midrule
MLP-5   & 4  & —              & 1.14\,M  & 4$\times$500\\
CNN-7   & 6  & every 2 layers & 1.70\,M  & 64, \, 2$\times$128, 3$\times$256\\
CNN-7W  & 6  & every 2 layers & 6.79\,M  & 128, 2$\times$256, 3$\times$512\\
CNN-12  & 11 & every 4 layers & 10.5\,M  & 64, \, 2$\times$128, 4$\times$256, 4$\times$512\\
\bottomrule
\end{tabular}
\end{table}

Architectures are denoted CNN-$X$ or MLP-$X$, where $X-1$ is the number of trainable feature layers and the final stage is the classification mechanism of the learning rule under evaluation. All trainable layers throughout this work (convolutional, fully connected, and the optional classifier head), use weight matrices without bias terms, so that, together with the parameter-free RMS normalization, every trainable parameter of the network is a weight that maps directly onto a differential device pair in the analog implementation. All evaluated networks are plain feedforward architectures, in line with prior forward-only learning studies. Residual connections would introduce cross-layer data paths whose activations must be buffered across layer boundaries and merged downstream, breaking the strictly layer-local data flow of the greedy scheme. Every convolutional layer applies a $3\times3$ convolution with stride 1 and same padding, initialized with the He-normal scheme, followed by RMS normalization and a ReLU activation. Spatial downsampling is performed exclusively by $2\times2$ max pooling with stride 2, inserted after every second convolutional layer (after every fourth layer for CNN-12, which reaches the same three spatial stages at greater depth). The convolutional stack is resolution-agnostic and is applied unchanged to every dataset. After the last convolutional layer, features are aggregated by global average pooling. BP and DFA require a fully connected softmax classifier appended after this aggregation, as both rules derive their error signal from a global network output. CwC, DCL, l-DCL, and CwSG require no classifier layer, since DCL and l-DCL read out class scores through their dictionary of embeddings, while CwC and CwSG obtain a class prediction in every layer from the per-class goodness. Table~\ref{tab:archs} lists the layer widths of each variant. The Wide variant scales all channel widths uniformly by a factor of 2, leaving depth, pooling, and every other convention unchanged.

MLP-5 follows the same conventions with fully connected layers. The input image is flattened and processed by four hidden layers of 500 units, each applying a bias-free dense projection initialized with the He-normal scheme, followed by the same parameter-free RMS normalization and ReLU activation; the classifier requirement is identical, with a classifier appended for BP and DFA only. For CwSG on fully connected layers, where no 2D spatial neighborhood exists, the spatial-gradient goodness is replaced by its 1D analog, computed from central differences between neighboring neurons within each class group.

\subsection*{Optimizers}\label{Methods:optimizers}
\subsubsection*{Stochastic Gradient Descent (SGD)}
To evaluate the learning algorithms, we implement three optimization schemes. As a baseline, full-precision training is performed using SGD, which could be applied to devices using a costly program-and-verify (PnV) scheme. It applies continuous updates $\Delta W$ proportional to the gradient $g_{ij} = \frac{\partial \mathcal{L}}{\partial W_{ij}}$, scaled by a learning-rate hyperparameter $\eta$, such that
\begin{equation}\label{eq:sgd}
    \Delta W_{ij} = -\eta g_{ij}.
\end{equation}

\subsubsection*{Sign-only Stochastic Gradient Descent (SignSGD)}
To model physical crossbar arrays where programming events are restricted to single, fixed-amplitude pulses (soft SET and RESET regimes), we constrain the updates using SignSGD. This optimizer discards the gradient magnitude and retains only its direction, yielding
\begin{equation}\label{eq:signsgd}
    \Delta W_{ij} = -\eta \operatorname{sign}(g_{ij}).
\end{equation}

\subsubsection*{Stochastic Sign-only Stochastic Gradient Descent (StoSignSGD)}
To bridge the performance gap for algorithms reliant on the gradient magnitude without violating the requirement for single-pulse physical updates, we use a stochastic StoSignSGD optimizer. This optimizer gates the fixed-amplitude pulses using a Bernoulli mask whose firing probability scales with the gradient magnitude, such that
\begin{equation}\label{eq:stosignsgd}
    \Delta W_{ij} = -\eta \operatorname{sign}(g_{ij}) b_{ij}, \quad \text{with} \quad b_{ij} \sim \mathrm{Bernoulli}\big(\operatorname{clip}(|g_{ij}|\theta, 0, 1)\big),
\end{equation}
where $\theta$ is a gating gain hyperparameter governing saturation. In the unsaturated regime ($|g_{ij}|\theta < 1$), the expected update recovers a step proportional to SGD, with effective learning rate $\eta\theta$:
\begin{equation}\label{eq:stosignsgd-exp}
    \mathbb{E}[\Delta W_{ij}] = -\eta \theta g_{ij}.
\end{equation}
StoSignSGD therefore encodes the gradient magnitude into the stochastic firing rate of constant-amplitude pulses.

\subsection*{Greedy training schedule}

All layer-wise rules train under the greedy schedule unless stated otherwise. Layers are trained strictly sequentially. Layer $\ell$ receives its full update budget of $N$ epochs, while the already-trained layers $1,\dots,\ell-1$ act as a frozen read-only feature extractor and the layers above $\ell$ remain at their random initialization. Because every objective is purely local, no forward pass beyond layer $\ell$ is required during its training phase, and no device outside array $\ell$ is programmed. The update budget is matched to the end-to-end baselines by construction, since $N$ epochs per layer under the greedy schedule and $N$ end-to-end epochs give every trainable parameter the same number of update opportunities. During greedy training, the test accuracy is probed at the layer currently being trained through its per-layer prediction readout. On hardware, this schedule implies that each crossbar is written during a single contiguous phase and only read afterward.

\subsection*{Batch storage}

Let $A_j$ denote the number of activations at layer $j$ for one sample, $A_0$ the number of inputs for the network, $m$ the batch size, and $L$ the network depth. Backpropagation retains the output of every layer for its backward pass, requiring
\begin{equation}
    M_{\mathrm{BP}}^{\mathrm{act}} = O\!\left(m\sum_{j=0}^{L}A_j\right).
\label{eq:mem_bp}
\end{equation}
Greedy training of layer $\ell$ instead retains only the input and output of the active layer, while the frozen feature extractor (layers $0$ to $\ell-1$) discards its activations as soon as the next layer is computed. This gives
\begin{equation}
    M_{\mathrm{local}}^{\mathrm{act}} = O\!\left(m\,(A_{\ell-1}+A_\ell)\right).
\label{eq:mem_local}
\end{equation}
Local activation storage is therefore bounded by the most demanding layer and independent of depth, whereas that of BP grows with depth $L$.

These buffers are sufficient to form the batched weight update, because the batch gradient of the active layer is the sum of its per-sample gradients. It can be computed either directly from the $m$ buffered inputs and local errors counted in Eq.~\ref{eq:mem_local}, or by processing the samples one at a time, summing their gradients into an accumulator with one entry per weight of the active layer. The latter replaces the batch activation buffer by a memory of size $O(P_\ell)$, with $P_\ell$ the number of active-layer weights. In both cases, no network-wide computation graph is ever stored.

\subsection*{Class-to-channel encoding}

For CwC and CwSG, the output channels of each convolutional layer (units of each hidden layer for MLP-5) are partitioned into $K$ disjoint groups, where $K$ is the number of classes, and the standard one-hot encoding maps each class to one group. This encoding requires the channel count $C_\ell$ of layer $\ell$ to be a multiple of $K$. When it is not, the $C_\ell \bmod K$ residual channels are excluded from the class mapping (for example, 4 of the 64 conv1 channels on CIFAR-10). They are however included in the normalization and gradient computation, and therefore receive non-zero updates. This one-hot construction breaks down entirely once the number of classes exceeds the width of a layer ($K > C_\ell$), since not every class can be assigned a channel, and it becomes wasteful whenever $K$ is large compared to $C_\ell$, as each class is then represented by very few channels.

\subsection*{Noise-injection strategy}

All three noise models perturb the weights multiplicatively with i.i.d. zero-mean Gaussian variables $\gamma \sim \mathcal{N}(0, \sigma_\gamma^2)$, applied element-wise, and differ only in which quantity carries the perturbation and when it is drawn. The two programming models (update and target noise) inject noise exactly once per optimizer step, at the moment the optimizer's update is committed to the stored weights, so a parameter that receives no update at a given step receives no noise either. 

Update noise, used for the SignSGD and StoSignSGD experiments, is implemented as $\widetilde{W} = W + \Delta W\,(1+\gamma)$, so the perturbation scales strictly with the magnitude of the applied update. Because sign-based optimizers apply fixed-magnitude steps $|\Delta W| = \eta$, the absolute noise injected per committed write has standard deviation $\eta\,\sigma_\gamma$ in weight units. We therefore sweep the learning-rate-normalized level $\sigma_{\mathrm{eff}} = \eta\,\sigma_\gamma$, implemented by dividing the injected perturbation by the tuned learning rate of each algorithm, so that a given $\sigma_{\mathrm{eff}}$ produces the same physical per-write perturbation for every learning rule even though the tuned $\eta$ of BP and CwSG differ by more than an order of magnitude. Without this normalization, the same nominal $\sigma_\gamma$ would inject absolute noise proportional to each rule's learning rate, and the comparison between algorithms would be confounded by their hyperparameters. Under StoSignSGD, the Bernoulli gate acts before the injection, so coordinates that emit no pulse at a given step receive neither an update nor noise, making the injected noise update-gated exactly as in a pulse-based write scheme.

Target noise, used for the SGD experiments, is implemented as $\widetilde{W} = (W + \Delta W)\,(1+\gamma)$, so the perturbation scales with the programmed target weight and maintains a constant signal-to-noise ratio with respect to the stored value, as achieved by PnV schemes. The swept level $\sigma_\gamma$ requires no learning-rate normalization, since the perturbation is anchored to the weight magnitude rather than to the update. The noise is applied to every weight written at the given step.

Readout noise is implemented as $\widetilde{W} = W\,(1+\gamma)$ and is purely transient, since the stored weights are never modified. A fresh, independent noise tensor is drawn for every input sample and for every physical access to the array, in both training and evaluation. Forward and backward passes constitute separate reads, so backpropagation error traverses an independently perturbed $\widetilde{W}^{\top}$, while the weight gradient is computed against the forward-pass realization. Read noise is applied to every weight-bearing operation, including all convolutional kernels and the dense readout.

\subsection*{Device characterization}

\begin{table}[h!]
\centering
\caption{\textbf{Programming and read conditions for the characterized memory arrays.} Word line (WL), bit line (BL), and source line (SL) voltages are provided alongside the exact pulse timing profile: rise time ($T_\text{rise}$), plateau duration ($T_\text{high}$), and fall time ($T_\text{fall}$). The soft programming operations (bold) constitute the strictly fixed-amplitude, unverified updates used for experimental measurements.}
\label{tab:dev_charac}
\begin{tabular}{llcccccc}
\toprule
\multirow{2}{*}{\textbf{Technology}} & \multirow{2}{*}{\textbf{Operation}} & \multicolumn{3}{c}{\textbf{Applied Voltages (V)}} & \multicolumn{3}{c}{\textbf{Pulse Timing}} \\
\cmidrule(lr){3-5} \cmidrule(lr){6-8}
 & & \textbf{WL} & \textbf{BL} & \textbf{SL} & $\boldsymbol{T_\text{rise}}$ \textbf{(ns)} & $\boldsymbol{T_\text{high}}$ & $\boldsymbol{T_\text{fall}}$ \textbf{(ns)} \\
\midrule
\textbf{Filamentary} & Forming & 2.8 & 2.4 & 0 & 100 & 10\,$\mu$s & 100 \\
\textbf{Memristors}  & SET & 2.5 & 2.0 & 0 & 100 & 50\,$\mu$s & 100 \\
(1{,}268 devices)    & RESET & 4.5 & 0 & 2.0 & 100 & 1\,$\mu$s & 100 \\
                     & \textbf{Soft RESET} & \textbf{4.5} & \textbf{0} & \textbf{0.9} & \textbf{100} & \textbf{600\,ns} & \textbf{100} \\
                     & Read & 4.5 & 0.2 & 0 & 100 & 5\,$\mu$s & 100 \\
\midrule
\textbf{Phase-Change} & RESET & 4.8 & 0.1 -- 4.9$^\dagger$ & 0 & 20 & 500\,ns & 20 \\
\textbf{Memory}       & \textbf{Soft SET} & \textbf{1.0} & \textbf{0.5} & \textbf{0} & \textbf{20} & \textbf{100\,ns} & \textbf{700} \\
(1{,}189 devices)     & Read & 4.8 & 0.4 & 0 & 10 & 1\,$\mu$s & 10 \\
\bottomrule
\end{tabular}
\par\smallskip
\parbox{\textwidth}{\raggedright\footnotesize $^\dagger$Swept from 0.1\,V to 4.9\,V in 0.2\,V increments.}
\end{table}

\subsubsection*{Filamentary memristors}

Filamentary memristor conductance trajectories were measured on a fabricated crossbar array \cite{harabi2023multimode}, a 128$\times$64 one-transistor-one-resistor (1T1R) architecture integrating hafnium oxide (HfO$_x$) memristive devices over a 130\,nm complementary metal-oxide-semiconductor (CMOS) front end. Each 300\,nm-diameter device features a stacked configuration consisting of titanium nitride (TiN), HfO$_x$, titanium (Ti), and a top TiN electrode. Source lines are shared by adjacent cells, and the array is accessed in direct analog mode for both reading and programming. With differential pairing, its 8,192 devices implement up to 4,096 synapses. Devices were formed with a fixed-amplitude scheme rather than a voltage ramp: identical pulses were applied repetitively (one to five shots) until conduction appeared, followed by three SET-RESET wake-up cycles that reinforce a thick, stable filament and initiate a progressive, approximately monotonic low-voltage RESET slope. The update scheme exploited in this work is not precise conductance targeting through iterative PnV loops, but a low-stress, state-agnostic update that gradually decrease conductance under identical sub-1\,V RESET pulses. The 0.9\,V amplitude balances the two failure modes of the regime, as lower amplitudes shrink the per-pulse decrement below the stochastic variability and degrade linearity, while higher amplitudes trigger premature filament dissolution, with single pulses dropping the conductance by more than 25\,$\mu$S on 20.3\% of devices at 1.0\,V, against 2.6\% at 0.9\,V and none at 0.8\,V. Beyond roughly 2,000 to 3,000 RESET pulses, the trajectories grow markedly more variable. The statistics used in the device-calibrated simulations are derived from 5,000 RESET pulses at 0.9\,V and 600\,ns duration (Table~\ref{tab:dev_charac}) applied to a cohort of 1,268 randomly selected devices across the array; the full optimization of this programming regime is described in ref~\cite{renaudineau2026}.

\subsubsection*{Phase-change memory}

Phase-change memory (PCM) conductance trajectories were measured on 1T1R cells integrated into a 4\,kb crossbar array, fabricated in the back-end-of-the-line (BEOL) of a 130\,nm CMOS process. The devices employ a 300\,nm ``Wall'' integration architecture \cite{Decaramet2024_PCM_CEA_JOEL}, but omit the germanium nitride (GeN) thermal confinement layer used in more highly scaled iterations of this technology. To establish a uniform high-resistance amorphous baseline, each device is first initialized using a staircase RESET sequence. This sequence applies 25 pulses at a constant word-line voltage of 4.8\,V while sweeping the bit-line voltage from 0.1\,V to 4.9\,V in 0.2\,V increments (Table~\ref{tab:dev_charac}), with a non-disturbing read operation after each step. Instead of targeting precise conductance values through PnV loops, the update scheme exploited in this work relies on state-agnostic, single-pulse updates that gradually increase conductance via progressive crystallization of the germanium-antimony-telluride (GST) active volume. Following initialization, devices are subjected to 10,000 identical soft SET pulses (word-line voltage $V_\text{WL} = 1.0$\,V, bit-line voltage $V_\text{BL} = 0.5$\,V). The extended 700\,ns fall time of these pulses facilitates gradual structural relaxation, driving a monotonic, cumulative increase in conductance. To trace the exact conductance evolution used in the network simulations, a read operation ($V_\text{BL} = 0.4$\,V) is interleaved between successive programming pulses. The device-calibrated simulations use these measured 10,000-pulse trajectories, acquired from a cohort of 1,189 randomly selected devices across the array.

\subsubsection*{Weight encoding}
To represent signed values, each weight is encoded using a differential conductance pair with a positive and a negative device such that

\begin{equation*}
W \;=\; s\,\left(G^{+}-G^{-}\right),   
\end{equation*}

where $s$ is a system-level gain  hyperparameter identical across all cells.

Because the network contains far more weights than the pools of measured phase-change and filamentary devices, the differential pairs are assigned by sampling without replacement. Each weight receives two devices drawn at random from the pool, and once too few devices remain to form a pair, the full pool is restored and reshuffled before the assignment continues, until every weight is mapped. Measured devices are therefore reused across the network, each an approximately equal number of times, while every differential pair is always formed from two distinct physical devices.

No initialization programming is carried out on the differential pair; the initial analog weights are determined by the state of the devices after a RESET operation (Table~\ref{tab:dev_charac}).

\subsection*{Data augmentation}
CIFAR-10 is augmented with a standard augmentation pipeline. Each training sample receives zero padding (4 pixels) followed by random cropping, and a random horizontal flip. The augmented inputs are then normalized before being fed to the network. The augmentation is applied at runtime and resampled independently at every epoch.

\subsection*{Peripheral converter resolution}

Every layer's matrix multiplication is driven and read at $b$ bits by DACs and ADCs. The remaining computations are not quantized by the converter models.

Drives are unipolar and convert the input of each layer; the input of the first layer is scaled to $[0,1]$. RMS normalization sets each position's channel vector to unit root-mean-square before the rectifier, so a fixed drive range serves each interface: $4$ for every hidden array, $1$ for the network input, and $2$ for the head, whose drive is a spatial mean of rectified activations. A drive spreads $2^{b}$ levels over its range $[0,B]$ with step $B/(2^{b}-1)$.

A readout digitizes the pre-activations, before normalization and ReLU, over $2^{b}-1$ levels on $[-B_\ell,+B_\ell]$ with step $\Delta_\ell = 2B_\ell/(2^{b}-2)$, rounding to the nearest code and clipping at $\pm B_\ell$. With $F_\ell$ denoting the number of products a column accumulates, $9\,C_{\text{in}}$ for a convolution and the last-layer channel count ($512$) for the head,
\begin{align*}
    B_\ell = C\,w_{\text{rail}}\,\sqrt{F_\ell} \ ,
    \qquad
    w_{\text{rail}} = \left| s \,(G_{\max}-G_{\min}) \right| \ ,
    \qquad
    C = 0.5.
\end{align*}

Both quantizers use an unclipped straight-through estimator: the backward pass treats each quantizer as the identity, assigning a gradient of one everywhere, including for saturation conversions. BP reads its error from the same arrays driven during the forward pass, which requires a second pair of bipolar converter banks: a drive presenting the error $\delta$ so that the array computes $W^{\top}\delta$, and its readout. Following the backward implementation of AIHWKit \cite{LeGallo2023AIHWKit}, the error is divided by its absolute maximum before the drive, and the digital result is multiplied back after the readout, so the drive spans $[-1,1]$ and the readout keeps the fixed range $B_\ell$ of the forward pass. Two BP configurations are defined. BP (FP32 bwd) is a numerical control in which the backward matrix products are evaluated without converter quantization. This isolates the effect of forward quantization and does not represent an implementation that eliminates the backward datapath. The second, more realistic configuration (BPq) quantizes the backward drive and the backward readout at the same $b$ bits as the forward pass, so each array carries drive and readout converter banks on both edges and the crossbar becomes bidirectional, with the additional peripheral circuitry this implies.

\end{document}